\documentclass[aps,prl,twocolumn,superscriptaddress,letterpaper,nobibnotes]{revtex4-2}

\usepackage{graphicx}
\usepackage{bm}
\usepackage{amssymb}
\usepackage{color}
\usepackage{amsmath}
\usepackage{physics}
\usepackage[colorlinks=true,linkcolor=blue,anchorcolor=black,
            citecolor=blue,filecolor=black,menucolor=black,
            runcolor=black,urlcolor=blue]{hyperref}

\renewcommand{\thefigure}{\arabic{figure}}
\makeatletter
\def\@caption@fignum@sep{\space}\let\@caption@fnum@sep\@caption@fignum@sep
\makeatother

\makeatletter
\def\@caption@fnum@sep{\space}
\makeatother

\renewcommand{\figurename}{Fig.}
\makeatletter
\renewcommand{\fnum@figure}{\footnotesize{\figurename~\thefigure}}
\makeatother

\begin{document}

	\title{Observation of localization reversal and harmonic generation in nonlinear non-Hermitian skin effect}

	\author{Junyao Wu}\thanks{These authors contributed equally to this work.}
	\affiliation{Interdisciplinary Center for Quantum Information, State Key Laboratory of Modern Optical Instrumentation, Zhejiang University, Hangzhou 310027, China}
	\affiliation{International Joint Innovation Center, The Electromagnetics Academy at Zhejiang University, Zhejiang University, Haining 314400, China}
	\affiliation{Key Lab. of Advanced Micro/Nano Electronic Devices \& Smart Systems of Zhejiang, Jinhua Institute of Zhejiang University, Zhejiang University, Jinhua 321099, China}
	\affiliation{Shaoxing Institute of Zhejiang University, Zhejiang University, Shaoxing 312000, China}

	\author{Rui-Chang Shen}\thanks{These authors contributed equally to this work.}
	\affiliation{Department of Physics, The Chinese University of Hong Kong, Shatin, Hong Kong SAR, China}

	\author{Li Zhang}
	\affiliation{Interdisciplinary Center for Quantum Information, State Key Laboratory of Modern Optical Instrumentation, Zhejiang University, Hangzhou 310027, China}
	\affiliation{International Joint Innovation Center, The Electromagnetics Academy at Zhejiang University, Zhejiang University, Haining 314400, China}
	\affiliation{Key Lab. of Advanced Micro/Nano Electronic Devices \& Smart Systems of Zhejiang, Jinhua Institute of Zhejiang University, Zhejiang University, Jinhua 321099, China}
	\affiliation{Shaoxing Institute of Zhejiang University, Zhejiang University, Shaoxing 312000, China}

	\author{Fujia Chen}
	\affiliation{Interdisciplinary Center for Quantum Information, State Key Laboratory of Modern Optical Instrumentation, Zhejiang University, Hangzhou 310027, China}
	\affiliation{International Joint Innovation Center, The Electromagnetics Academy at Zhejiang University, Zhejiang University, Haining 314400, China}
	\affiliation{Key Lab. of Advanced Micro/Nano Electronic Devices \& Smart Systems of Zhejiang, Jinhua Institute of Zhejiang University, Zhejiang University, Jinhua 321099, China}
	\affiliation{Shaoxing Institute of Zhejiang University, Zhejiang University, Shaoxing 312000, China}

	\author{Bingbing Wang}
	\affiliation{Department of Physics, The Chinese University of Hong Kong, Shatin, Hong Kong SAR, China}

	\author{Hongsheng Chen}
	\affiliation{Interdisciplinary Center for Quantum Information, State Key Laboratory of Modern Optical Instrumentation, Zhejiang University, Hangzhou 310027, China}
	\affiliation{International Joint Innovation Center, The Electromagnetics Academy at Zhejiang University, Zhejiang University, Haining 314400, China}
	\affiliation{Key Lab. of Advanced Micro/Nano Electronic Devices \& Smart Systems of Zhejiang, Jinhua Institute of Zhejiang University, Zhejiang University, Jinhua 321099, China}
	\affiliation{Shaoxing Institute of Zhejiang University, Zhejiang University, Shaoxing 312000, China}

	\author{Yihao Yang}\email[\# Corresponding author: ]{yangyihao@zju.edu.cn}
	\affiliation{Interdisciplinary Center for Quantum Information, State Key Laboratory of Modern Optical Instrumentation, Zhejiang University, Hangzhou 310027, China}
	\affiliation{International Joint Innovation Center, The Electromagnetics Academy at Zhejiang University, Zhejiang University, Haining 314400, China}
	\affiliation{Key Lab. of Advanced Micro/Nano Electronic Devices \& Smart Systems of Zhejiang, Jinhua Institute of Zhejiang University, Zhejiang University, Jinhua 321099, China}
	\affiliation{Shaoxing Institute of Zhejiang University, Zhejiang University, Shaoxing 312000, China}

	\author{Haoran Xue}\email[\# Corresponding author: ]{haoranxue@cuhk.edu.hk}
	\affiliation{Department of Physics, The Chinese University of Hong Kong, Shatin, Hong Kong SAR, China}
	\affiliation{State Key Laboratory of Quantum Information Technologies and Materials, The Chinese University of Hong Kong, Shatin, Hong Kong SAR, China}

	\begin{abstract}
		The interplay between band topology and material nonlinearity gives rise to a variety of novel phenomena, such as topological solitons and nonlinearity-induced topological phase transitions. However, most previous studies fall within the Hermitian regime, leaving the impact of nonlinearity on non-Hermitian topology seldom explored. Here, we investigate the effects of nonlinearity on the non-Hermitian skin effect, a well-known non-Hermitian phenomenon induced by the point-gap topology unique to non-Hermitian systems. We discover a nonlinearity-induced point-gap topological phase transition accompanied by a reversal of the skin mode localization. This phenomenon is experimentally demonstrated in a nonlinear microwave metamaterial, where electromagnetic waves are localized around one end of the sample under a low-intensity pump, whereas at a high-intensity pump, the waves are localized around the other end. We also observe third harmonic generation signal induced by the skin modes, whose spatial distribution consistently shows the localization reversal. {Furthermore, we extend our setup to build a nonlinear non-Hermitian Su–Schrieffer–Heeger model with both nonlinearity-induced point-gap and line-gap phase transitions.} Our results open a new route towards nonlinear topological physics in non-Hermitian systems and are promising for reconfigurable topological wave manipulation and frequency conversion.
	\end{abstract}
	\maketitle

	\noindent\textbf{Introduction}\\
	Topological photonics, a field where the concept of band topology developed in condensed matter is utilized to design photonic structures, offers a promising route towards robust manipulation of electromagnetic waves~\cite{lu2014topological, ozawa2019topological}. Since the realization of photonic Chern insulators at microwave frequencies~\cite{wang2008reflection, wang2009observation}, various topological phases have been successfully implemented in photonics, including the quantum spin Hall phase~\cite{khanikaev2013photonic, chen2014experimental, wu2015scheme}, valley Hall phase~\cite{ma2016all, dong2017valley, gao2018topologically}, Floquet topological phase~\cite{rechtsman2013photonic, liang2013optical, gao2016probing}, among others. A key factor behind these rapid developments is that the source-free Maxwell equation can be cast into a linear eigenproblem, whose eigenmodes can be used to define topology using standard topological band theory. However, there are two noticeable exceptions, i.e., non-Hermitian and nonlinear systems, where conventional topological band theory is not applicable. The study of these systems has led to a new revolution in topological photonics, together with various novel topological phenomena~\cite{bergholtz2021exceptional, ding2022non, smirnova2020nonlinear}.

	Non-Hermiticity, resulting from nonconservation of energy, is ubiquitous in photonic systems due to material loss and/or gain. When it is taken into consideration, the topological classification and even the notion of topology will be significantly altered, leading to many topological phenomena without Hermitian counterparts~\cite{gong2018topological, kawabata2019symmetry, zhou2019periodic}. One prominent example is the non-Hermitian skin effect (NHSE), in which an extensive number of eigenmodes are localized at an open boundary due to the nontrivial point-gap topology under the periodic boundary condition (PBC)~\cite{yao2018edge, kunst2018biorthogonal, martinez2018non, borgnia2020non, okuma2020topological, zhang2020correspondence}. The NHSE has been realized in photonics using different platforms~\cite{xiao2020non, weidemann2020topological, liu2022complex, sun2024photonic, liu2024localization,manna2023inner}, with potential applications in lasing and sensing~\cite{longhi2018non, zhu2022anomalous, gao2023two, budich2020non, mcdonald2020exponentially, deng2024ultrasensitive}.

	Meanwhile, nonlinearity has also been shown to have a significant impact on topological photonic systems. While proper definitions of band topology in nonlinear systems are still an ongoing topic, several interesting phenomena from the interplay between nonlinearity and band topology have already been discovered, such as modifications of topological modes from nonlinearity (e.g., to form solitons)~\cite{lumer2013self, dobrykh2018nonlinear, mukherjee2020observation, sahin2025protected, kirsch2021nonlinear, hu2021nonlinear, jurgensen2021quantized}, frequency conversion using topological modes~\cite{kruk2019nonlinear, smirnova2019third, wang2019topologically, lan2020nonlinear, you2020four} and topological phase transitions driven by nonlinearity~\cite{hadad2016self, leykam2016edge, leykam2018reconfigurable, maczewsky2020nonlinearity,sahin2025topolectrical}. In particular, nonlinearity-induced topological phase transitions, which go beyond the straightforward nonlinearity-topological mode interaction picture, suggest a deep and subtle interplay between band topology and nonlinearity and offer an appealing method to control topological propagation by tuning input power instead of system parameters.

	\begin{figure*}
		\centering
		\includegraphics[width =  0.9\textwidth]{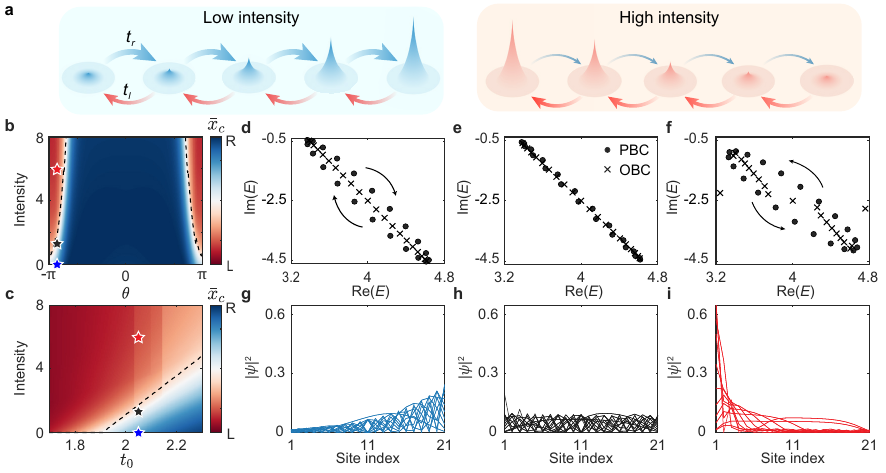}%
		\caption{{Nonlinear Hatano-Nelson model.} \textbf{a} Schematic of the nonlinear Hatano-Nelson model and nonlinearity-induced reversal of skin localization. Here, the $t_{{r}({l})}$ represents the rightward (leftward) coupling, which is nonlinear (linear). \textbf{b},\textbf{c} Plot of the average position of all OBC eigenvectors as functions of intensity $I$ and $\theta$ (and $t_0$),\, with a fixed $t_0=2.05$ ($\theta=-0.9\pi$).  The black dashed lines denote the point-gap closing points obtained from PBC spectra.  \textbf{d-f} Plots of eigenvalues on the complex plane under PBC (dots) and OBC (crosses) for the three intensity values highlighted by the blue (d), black (e), and red (f) star markers in \textbf{b} and \textbf{c}. \textbf{g-i} The corresponding OBC eigenvectors for \textbf{d}-\textbf{f}. In all calculations, we take $f_0=-2.5i$, $t_{l}=1$, $t_{\infty}=0.2$ and $t_{c}=1$.
		}
		\label{fig1}
	\end{figure*}

	While previous studies have revealed fruitful topological physics from non-Hermitian or nonlinear topological systems separately, little is known about systems that are simultaneously non-Hermitian and nonlinear~\cite{xia2021nonlinear, dai2024non}. As non-Hermitian systems can host point-gap topological phases that are absent in Hermitian ones, new physics is expected to emerge when they are enriched by nonlinearity. Recently, considerable efforts have been put into studying the impact of nonlinearity on the NHSE, showing that the skin modes can be significantly tuned by nonlinearity~\cite{yuce2021nonlinear, ezawa2022dynamical, jiang2023nonlinear, many2024skin, ghaemi2024class, yuce2025nonlinear, wang2024nonlinear, lo2024switchable, padlewski2025observation}. Particularly, previous studies in low-frequency systems have shown that nonlinearity can lead to the opening of a point gap, and thus, the emergence of skin modes~\cite{lo2024switchable, padlewski2025observation}.  However, whether nonlinearity can induce a topological phase transition between different nontrivial point-gap phases remains unexplored.

    In this work, we theoretically and experimentally discover that nonlinearity can induce a topological phase transition between two nontrivial point-gap topological phases in an electromagnetic non-Hermitian metamaterial at gigahertz frequencies.   {This phase transition leads to a localization reversal phenomenon: the modes originally localize at the right boundary are relocated to the left boundary.} Specifically, we introduce a nonlinear Hatano-Nelson model with saturable nonreciprocal couplings, which is known as the minimal model for the NHSE in the linear limit~\cite{hatano1996localization}. In our nonlinear case, we find that the direction of the NHSE, or equivalently, the point-gap topology, can be solely controlled by nonlinearity. Consequently, the field localization from the NHSE can be reversed by simply increasing the input power (see Fig.~\ref{fig1}a). This phenomenon is experimentally observed in a microwave metamaterial through port-resolved voltage measurements and near-field scanning of the electric field distribution. We also demonstrate for the first time the harmonic generation of the skin modes at high pump strength, and successfully capture the localized fields at the harmonic frequency that resemble the skin localization.   {Moreover, we construct a nonlinear non-Hermitian Su–Schrieffer–Heeger (SSH) model~\cite{su1979solitons} and discover that the point-gap and line-gap topology can be controlled by nonlinearity simultaneously. These findings extend the study of nonlinear topological phenomena to the point-gap regime.}

\noindent\textbf{Results}\\
\noindent\textit{Nonlinear Hatano-Nelson model.} Consider a   {classical} nonlinear Hatano-Nelson model as depicted in  Fig.~\ref{fig1}a, which generically describes an array of   {$N$ coupled photonic resonators} with linear reciprocal couplings and nonlinear nonreciprocal couplings.   {The system is governed by the coupled-mode equations:}
  {
\begin{eqnarray}\label{eq01}
\begin{aligned}
i \frac{d \Psi_n}{d t} = f_0 \Psi_n + t_{r,n} \Psi_{n-1} +t_{l,n} \Psi_{n+1},
\end{aligned}
\end{eqnarray}
}
 {where $n$ is the site index, $\Psi_n$ is the complex field amplitude at the $n$-th site}, $f_0$ is the resonant frequency, and $t_{{l}({r}), n}$ denotes the leftward (rightward) nearest-neighbor coupling. Here, the leftward coupling $t_{{l},n}=\kappa_{1, n}$ while the rightward coupling $t_{{r}, n}=\kappa_{1, n}+\tilde \kappa _{2, n}$, where $\kappa_{1, n}$ is a conventional linear and reciprocal coupling (e.g., an evanescent coupling) and $\tilde \kappa _{2, n}$ is a specially engineered nonlinear coupling that only exists in the rightward hopping process. A concrete realization of $\tilde \kappa _{2, n}$ will be introduced later. Since $\kappa_{1,n}$ is uniform across all sites, we write it as $\kappa_1$ hereafter. In the theoretical model, we set $\kappa_1 = 1$ as the normalization unit, while in the experimental sections $\kappa_1$ takes its physical value retrieved from measurements. The nonlinear nonreciprocal
coupling, which depends on the field intensity, is given by
\begin{eqnarray}\label{eq02}
\begin{aligned}
	\tilde \kappa _{2,n}(I_{n})=\left( \frac{t_0 - t_\infty}{1 + I_{n}/ t_c} + t_\infty \right) e^{i\theta} .
\end{aligned}
\end{eqnarray}

Here,   { $I_{n}=|\Psi_n|^2$} is the field intensity at the $n$-th site,  $t_{0}$ and $t_{\infty}$ correspond to the coupling strengths at zero and infinite intensity, respectively, $t_\text{c}$ governs how coupling varies with intensity, and $\theta$ is the phase of the coupling. Note that, due to the unidirectional nature of this coupling, its strength is only affected by the field intensity on the $n$-th site but not the $(n+1)$-th site. This modeling is also consistent with our experimental implementation, as we will illustrate later. In the linear Hatano-Nelson model, the localization direction of the skin modes is simply determined by the relative strength between the leftward and rightward couplings. Hence, in our case, as only one of them is nonlinear, it is intuitive to expect an intensity-driven reversal of the NHSE.

To investigate the NHSE in this model, {we search for time-harmonic solutions of the form $\Psi_n(t)=\psi_ne^{-iEt}$ to Eq.~\eqref{eq01}. This leads to the nonlinear eigenproblem $ H\psi=E\psi$, where $E$ is the eigenvalue, $\psi=(\psi_1,\psi_2,\ldots,\psi_N)^T$ is the eigenvector, and $H$ is the Hamiltonian containing the on-site and coupling terms (see Methods for more numerical details).} To account for the nonuniform coupling distribution induced by intensity fluctuation in space, we adopt a chain with $N=21$ sites and connect the first and last sites to form a closed loop to compute the PBC spectrum. To characterize the collective localization behavior of the eigenmodes, we calculate the average position of all modes under open boundary conditions (OBCs), defined as

\begin{equation}
\bar{x}_c=\langle\sum\limits_{j=1}^N|\psi_{j}|^2\frac{(j-(N+1)/2)}{(N-1)/2}\rangle,	\label{eq03}
\end{equation}
where $\psi_{j}$ is the amplitude of the eigenmode at site $j$, $\langle\cdot\rangle$ denotes the average over all eigenmodes, and $\bar x_c \in [-1, 1]$. A positive (negative) value of $\bar x_c$ indicates that the mode is localized at the right (left) end of the lattice.

Figure~\ref{fig1}b shows the calculated $\bar x_c$ as functions of the coupling angle $\theta$ and the total intensity     {$I=\sum_{n} |\psi_n|^2$}. The topological phase transition points of the NHSE are captured by $\bar x_c=0$, separating modes localized at opposite ends. Meanwhile, the transition can also be judged from the closing of the point-gap under PBC, as denoted by the black dashed lines, which coincide well with the $\bar x_c=0$ boundaries of the OBC modes. Notably, the intensity-driven reversal of the NHSE occurs near $\theta=\pm\pi$, where the interference between reciprocal and nonreciprocal couplings is significant. Whereas at $\theta=0$, $\kappa_1$ and $\tilde \kappa_2$ have the same sign regardless of the mode intensity, resulting in an absence of topological phase transition. In Fig.~\ref{fig1}c, we plot the distribution of  $\bar x_c$ in the $I-t_0$ plane for a fixed $\theta=-0.9\pi$, where a reversal of the NHSE is also found.   {A threshold, given by $|\kappa_{1}+\tilde \kappa _{2, n}(I_n\rightarrow 0)|>\kappa_1$, is required, such that rightward couplings are dominant at zero intensity.  In addition, the high-intensity coupling needs to satisfy $|\kappa_{1}+\tilde \kappa _{2, n}(I_n\rightarrow \infty)|<\kappa_1$ to make the reversal happen.}

	To see in detail the reversal process, we pick up three points in the phase diagrams (see the markers in Fig.~\ref{fig1}b,c) and plot the corresponding PBC and OBC eigenvalues and the OBC eigenmodes. As shown in Fig.~\ref{fig1}d-i, the modes are localized at opposite ends before and after the phase transition.  {A complementary winding-number characterization of the PBC Hamiltonian is provided in Supplementary Note 1.} Furthermore, the eigenvalues under OBC fall inside the loops of the PBC spectra, consistent with the feature of the NHSE.    All these numerical results suggest a point-gap topological phase transition induced by nonlinearity.

\begin{figure}[b]
\centering
\includegraphics[width=\columnwidth]{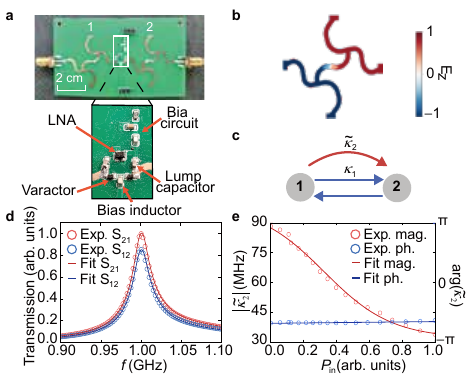}
\caption{{Implementation of microwave nonlinear nonreciprocal couplings.} \textbf{a} Photograph of a sample consisting of two coupled microwave resonators (labeled 1 and 2). The lower panel shows the details of the coupling region. \textbf{b} Simulated eigenmode profile of a single microwave resonator.  \textbf{c}  Simplified tight-binding model for the setup in \textbf{a}.  \textbf{d} Experimentally measured (circles) and numerically calculated (curves) response spectra for the setup in \textbf{a}. {The transmission is normalized to maximum linear response of $S_{21}$.} \textbf{e} Experimentally measured (circles) and numerically calculated (curves) magnitude and phase as a function of input power  {$P_{\text{in}}$}  at 1.056 GHz. The source and probe are at resonators 1 and 2, respectively. }
\label{fig2}
\end{figure}

\begin{figure*}
\centering
\includegraphics[width=\textwidth]{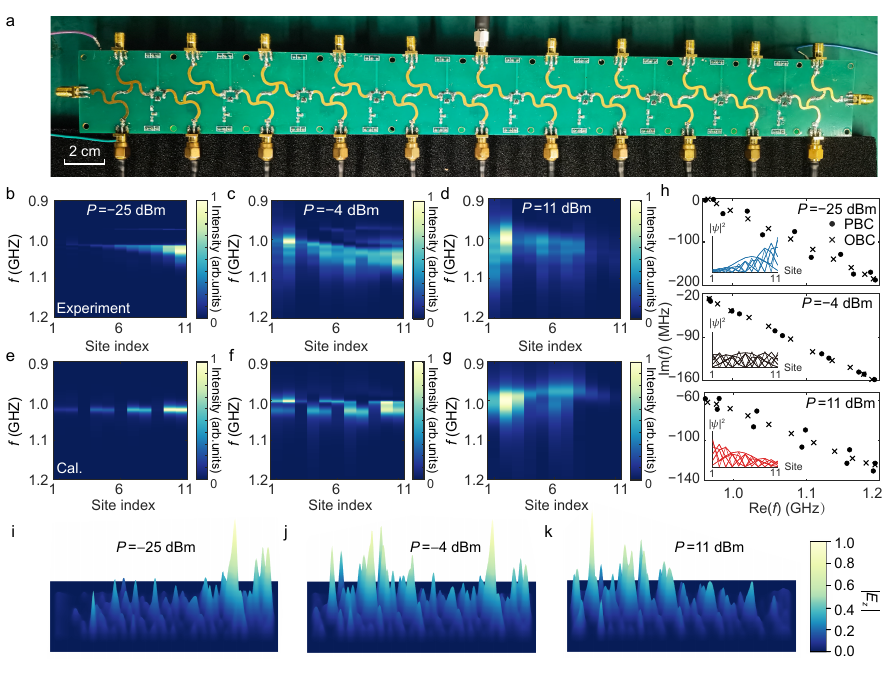}
\caption{{Observation of nonlinearity-induced reversal of the NHSE via port measurements.} \textbf{a} Photograph of a sample composed of 11 identical resonators. The source is connected to the lower ports, whereas the probe is affixed to the upper ports. \textbf{b}-\textbf{d} Measured field distributions at $P_{\text{in}} = -25$ dBm (\textbf{b}), $P_{\text{in}} = -4$ dBm (\textbf{c}), and $P_{\text{in}} = 11$ dBm (\textbf{d}), each normalized to its maximum response at the corresponding input power. \textbf{e}-\textbf{g} Simulated field distributions at $P_{\text{in}} = -25$ dBm (\textbf{e}), $P_{\text{in}} = -4$ dBm (\textbf{f}), and $P_{\text{in}} = 11$ dBm (\textbf{g}), likewise normalized at each power level. \textbf{h} Calculated PBC (circles) and OBC (crosses) eigenfrequencies and eigenmodes (insets) for the three cases in \textbf{b}-\textbf{d}. \textbf{i}-\textbf{k} Experimentally measured electric field distributions at 1.013 GHz ($|E_z|$ component) with $P_{\text{in}} = -25$ dBm (\textbf{i}), $P_{\text{in}} = -4$ dBm (\textbf{j}), and $P_{\text{in}} = 11$ dBm (\textbf{k}), each normalized to the maximum field amplitude at the corresponding power.}
\label{fig3}
\end{figure*}

\noindent\textit{Implementation of nonlinear nonreciprocal coupling at microwave frequencies.} To realize the theoretical model, we first propose a design to realize the nonlinear non-Hermitian coupling using two coupled microwave resonators. As illustrated in Fig.~\ref{fig2}a, the system consists of two identical resonators (labeled 1 and 2) made of copper strips on a FR4 substrate, with each supporting a dipolar mode at 1.2 GHz~\cite{peterson2018quantized} (see Fig.~\ref{fig2}b for the mode profile and {Supplementary Note 4 for more sample details}). The coupling between them is enabled in two ways. Firstly, the two resonators are connected by series varactors, which provide a reciprocal coupling $\kappa_1$ (see the lower panel of Fig.~\ref{fig2}a). This coupling can be adjusted by tuning the DC bias voltage (see Supplementary Note 3). Secondly, a unidirectional coupling, denoted by ${\tilde \kappa _2}$, is introduced by an RF amplifier circuit equipped with a low-noise amplifier chip (LNA) and a corresponding bias structure, as depicted in the lower panel of Fig.~\ref{fig2}a. The microwave signal from resonator 1 traverses the lumped capacitor and is detected by the LNA.   {Then it is unidirectionally amplified and coupled to resonator 2 via the other lumped capacitor. The amplitude gain and phase shift introduced by the LNA are captured by the complex coefficient  ${\tilde \kappa _2}$. At low input power, the LNA operates in the linear regime, and ${\tilde \kappa _2}$ remains effectively constant. As the input power increases, nonlinear gain compression occurs, where the fundamental amplification is reduced and waveform distortion emerges, leading to a power-dependent complex coupling coefficient. In this way, a nonlinear nonreciprocal coupling is realized in the present setup. In practice, the experimentally accessible parameter window of the coupling is limited by component tolerances, amplifier nonlinear operating range, and measurement noise, which may induce resonator detuning and introduce additional effective gain or loss. However, by choosing high-quality LNA and conducting careful calibration, the desired parameter values can be reached.}

Under the dipolar mode basis, the Hamiltonian of this system can be written as (see Fig.~\ref{fig2}c and  Supplementary Note 3 for the circuit-based derivation)
{
\begin{equation}
H_{\text{eff}}=\left(\begin{array}{cc}
f_0-i \gamma_0 -i\gamma_1& \kappa_1 \\
\kappa_1+\tilde{\kappa}_2 & f_0-i \gamma_0-i\gamma_1
\end{array}\right)
\label{Eq4}
\end{equation}
}
where $f_0$ is the resonance frequency of the
individual resonators, $\kappa_1$ and $\tilde{\kappa}_2$
are the reciprocal and nonreciprocal couplings. $\gamma_0$ arises from the
intrinsic loss, including the conductor loss, dielectric
loss, and lumped resistor loss, and $\gamma_1$ is the
external decay rate. Next, we measure the transmission spectra $|S_{21}|$ and $|S_{12}|$ under different input powers to validate the modeling and retrieve the model parameters. The measured spectra at input power $P_\mathrm{in}=-25$ dBm are shown in Fig.~\ref{fig2}d, which are not identical when the excitation and source positions are exchanged, indicating the presence of nonreciprocity.   {Using a Green's function method, we can establish the relationship between the transmission spectra and the tight-binding coefficients, allowing us to extract model parameters from experimental transmission data (see Supplementary Note 2).} Figure~\ref{fig2}e plots the retrieved magnitude and phase of  $\tilde \kappa_2$ against the input power, which fit well with the nonlinear coupling model (i.e., Eq.~\eqref{eq02}). Note that the phase of $\tilde \kappa_2$ is carefully engineered to ensure the reversal of the NHSE (see Fig.~\ref{fig1}b).

\noindent\textit{Reversal of the NHSE driven by nonlinearity.} We expand our system to a one-dimensional crystal consisting of 11 microwave resonators, as shown in Fig.~\ref{fig3}a, with the coupling configuration between two neighboring resonators the same as the one in Fig.~\ref{fig2}a.   {Note that we use fewer resonators ($N=11$) to facilitate the experimental implementation, and the localization reversal phenomenon persists under the size change of the system.} To excite the skin modes, a power splitter uniformly divides a source with tunable power into 11 distinct segments, which are then simultaneously introduced into the resonators through the lower SMA ports (see Methods for more experimental details). Such an excitation averages the initial energy of the system and mitigates the effect of input ports. We use two methods to obtain the field distribution in the sample, as described below.

	We first measure the transmission spectrum at each resonator through the other SMA ports as the data for each lattice site. With a low input power ($P_\mathrm{in}=-25$ dBm), the rightward coupling takes the lead and a skin localization around the right end of the sample is experimentally observed (Fig.~\ref{fig3}b), which is also consistent with the numerical eigenmodes and field distribution obtained using retrieved tight-binding parameters (Fig.~\ref{fig3}e,h). With increased power, the NHSE gradually becomes weaker and the field becomes almost delocalized at around $P_\mathrm{in}=-4$ dBm, as can be seen in Fig.~\ref{fig3}c,f,h. Furthermore, when the input power exceeds this transition value ($P_\mathrm{in}=11$ dBm), a reversal of the NHSE, highlighted by a field localization around the left end of the sample, occurs as predicted (Fig.~\ref{fig3}d,g,h).

	\begin{figure}[t]
		\centering
		\includegraphics[width=\columnwidth]{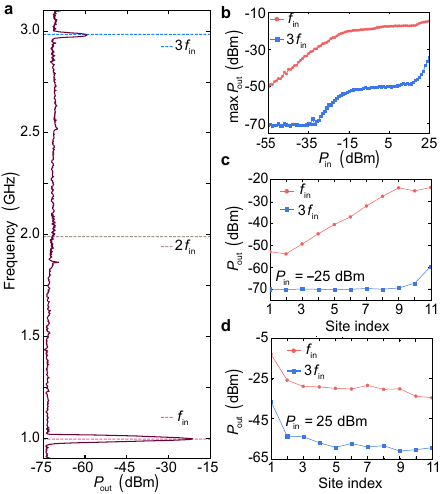}
	\caption{{Observation of harmonic signals produced by the skin modes.} \textbf{a} Measured output spectrum at site 11 under input power $P_{\text{in}} = -25$ dBm, {driven at $f_{\text{in}} = 0.994$ GHz (slightly shifted from 1.013 GHz due to the change in the measurement setup)}.  \textbf{b} Maximum output power among all sites at fundamental ($f_{\text{in}}$) and third ($3f_{\text{in}}$) harmonics as a function of increasing input power $P_{\text{in}}$. \textbf{c}, \textbf{d} Field localization behavior of fundamental and third harmonics, measured at $P_{\text{in}} = -25$ dBm and $P_{\text{in}} = 25$ dBm, {respectively}.}
		\label{fig4}
	\end{figure}

Additionally, we map out the electric field distribution (out-of-plane component) near the surface of the sample to more precisely capture the NHSE. The experimental results at 1.013 GHz for the three input power values are shown in Fig.~\ref{fig3}i-k. Despite now being two-dimensional field plots, the key characteristics (i.e., the localization behaviors) are the same as those revealed by the one-dimensional field plots given in Fig.~\ref{fig3}b-d using port measurements. Specifically, from low input power to high input power, we see a reversal of the NHSE from the right to the left end of the sample.  {The associated localization behavior is further quantified in Supplementary Note 7 through the localization length.}

\begin{figure*}[t]
\centering
\includegraphics[width=0.8\textwidth]{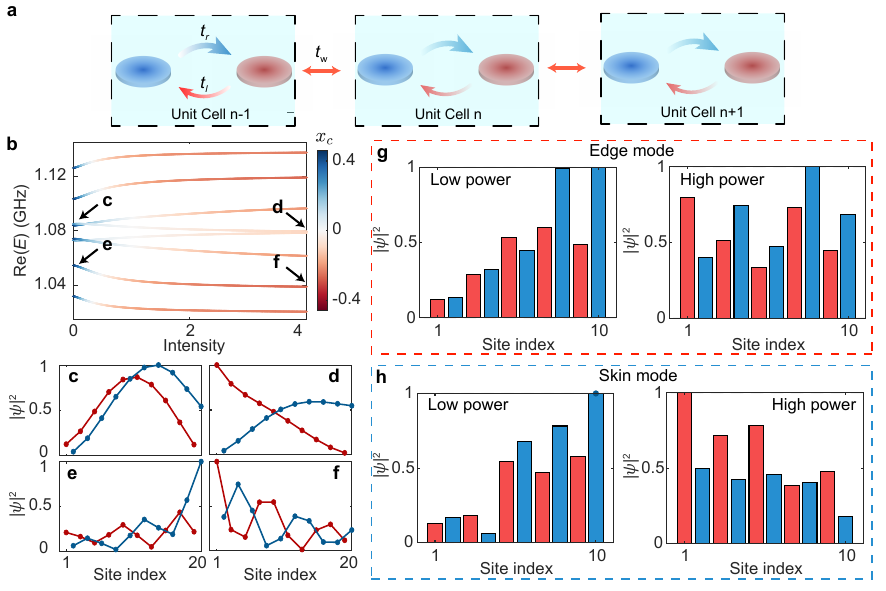}
\caption{{{Nonlinear nonreciprocal SSH model.} \textbf{a} Schematic of the nonlinear SSH model. \textbf{b} Evolution of the real part of eigenvalues with respect to the total intensity $I$. \textbf{c-f} The evolution of the eigenmodes, with their corresponding eigenvalues and intensities labeled in \textbf{b}. Red and blue curves correspond to the odd and even sites, respectively. \textbf{g}, \textbf{h}  Experimentally measured field distributions obtained by exciting the system with an input field whose frequency and phase correspond to the edge mode (\textbf{g}), and the skin mode (\textbf{h}).
The parameters used in the calculations are $t_l=t_w=66$ MHz, $t_0=98.9$ MHz, $t_\infty=32$ MHz, $\theta=-2.13$, and $t_c=0.022$.
}}
\label{fig5}
\end{figure*}
  {The nonlinear coupling strength sets how sensitively localization responds to increasing input power, while its phase determines whether this change preserves or reverses the original localization direction. A nonlinear contribution with the same sign as the intrinsic nonreciprocal bias maintains the localization direction, whereas an opposite sign enables reversal. Larger systems or higher-dimensional geometries can further enhance this behavior by allowing stronger energy accumulation under nonreciprocal transport.  Its robustness against disorder is further examined in Supplementary Note 6. }

\noindent\textit{Harmonic generation produced by skin modes.} Next, we investigate harmonic generation of the skin modes, which has not been  discussed in previous literature. For this purpose, we supply a sinusoidal continuous-wave input voltage signal with tunable frequency $f_{\rm in}$ and power $P_{\rm in}$ at site 1, then collect the spectrum of all sites. {The single-port excitation scheme is adopted intentionally, because if multiple ports are driven simultaneously, each excitation channel may introduce weak parasitic higher-harmonic components from the source and feeding network, thereby leading to unwanted harmonic contributions in the measured signal.} A typical measurement result at port 11 is shown in Fig.~\ref{fig4}a, for ${P_{\rm in}} =  - 25~{\rm{ dBm}}$ and input frequency $f_{\rm in}$ ranging from 0.95 to 1.05 GHz (i.e., at the frequencies of the skin modes). A clear peak at the third harmonic frequency ($3f_{\rm in}$) is observed, as indicated by the blue dashed line. The harmonic signal stems from the saturable nonlinearity of the LNAs, and even harmonics are suppressed due to the symmetry of the RF chip (see Supplementary Note 5). The third harmonic component appears only after the power of the fundamental mode exceeds a threshold level, and then grows concurrently with the fundamental as the input power increases, as demonstrated in Fig.~\ref{fig4}b. More interestingly, the field distributions at the third harmonic frequency inherit the localization property of the skin modes at the fundamental frequency. This is shown in Fig.~\ref{fig4}c and Fig.~\ref{fig4}d, where the localization direction of the third harmonic signals reverses following the nonlinearity-induced topological phase transition at the fundamental frequency when the input power increases from -25 dBm to 25 dBm. These observations indicate that we can control the spatial distribution of the harmonic fields using a tunable NHSE at fundamental frequencies.

\noindent\textit{Nonlinear nonreciprocal SSH model.} Finally, we extend our discussion to a nonlinear nonreciprocal SSH, whose linear version exhibits a series of novel phenomena from the interplay between point-gap and line-gap topologies, such as the breakdown of bulk-boundary correspondence and morphing of topological edge modes~\cite{yao2018edge, wang2022non}. Specifically, the point-gap topological phase transition manifests as the reversal of the NHSE, while the line-gap phase transition governs the emergence of topological edge states. As shown in Fig.~\ref{fig5}a, a rightward nonlinear coupling $t_r$ (following Eq.~\eqref{eq02}) is introduced to the intracell coupling. All other couplings remain linear. The eigenvalues and eigenmodes as functions of total intensity are plotted in Fig.~\ref{fig5}b–f for a finite chain under OBC. With initial parameter conditions $|t_r|\cdot|t_l| > |t_w|^2$ and $|t_r| > |t_l|$,    the system is topologically trivial at low intensities, and the skin modes are localized around the right end (Fig.~\ref{fig5}c,e).  As the total intensity increases, the real parts of the eigenvalues of two modes gradually approach each other, corresponding to a line-gap phase transition. However, due to inevitable mode coupling in finite-length chains, these two modes are not exactly degenerate~\cite{hadad2016self}. Meanwhile, a point-gap phase transition also occurs, as can be seen from the blue-to-red color change in Fig.~\ref{fig5}b and the eigenmodes plots in Fig.~\ref{fig5}e,f.

To further distinguish between edge modes and skin modes, we label the odd and even lattice sites with different colors. When the coupling strengths satisfy $|\ln(|t_r|/|t_l|)| < 2$, the edge modes exhibit opposite localization directions on odd and even sites, whereas the skin modes do not possess this property~\cite{zhu2021delocalization}. In our model, the nonlinear coupling strength ensures that this condition holds for all sites under arbitrary system intensities. As shown in Fig.~\ref{fig5}d,f, at high intensities, the edge modes exhibit opposite localization directions on odd and even sites, while the skin modes localize in the same direction. This confirms the emergence of edge modes. In the experiment, we clearly observe that, as the input power increased, the field distribution on the odd and even sites at the frequency corresponding to the edge mode changes from same-side localization to opposite-side localization (Fig.~\ref{fig5}g), accompanied by a reversal of the NHSE (Fig.~\ref{fig5}h). This indicates the simultaneous realization of both point-gap and line-gap phase transitions in the nonlinear SSH model.

\noindent\textbf{Discussion}\\
In summary, we have proposed a nonlinearity-induced topological phase transition of the unique non-Hermitian point-gap topology and experimentally observed the reversal of the NHSE driven by input power and the harmonic generation linked to the NHSE.  On the fundamental level, our results demonstrate that nonlinearity can drive point-gap topological phase transitions in non-Hermitian systems, enabling the reversal of the NHSE
through input power rather than structural modification. The harmonic signals generated by the skin modes further inherit this reversal, revealing an intrinsic connection between nonlinear frequency conversion and non-Hermitian topology. Practically, our scheme to realize the nonreciprocal coupling, which is much more compact and controllable compared to previous designs that require feedback controls~\cite{wang2024nonlinear,padlewski2025observation}, is promising for reconfigurable and nonreciprocal manipulation of electromagnetic waves. Meanwhile, the successful generation of harmonic signals may open a new route for frequency conversion and generation using novel non-Hermitian mechanisms.

      {The key element in our system, i.e., amplifier-based nonreciprocal coupling combined with saturable nonlinearity, is not restricted to this specific implementation and is in principle applicable to other active wave platforms. Moreover, the current four-arm H-shaped resonator design naturally extends to two-dimensional networks, as each arm offers an independent coupling channel, facilitating planar connectivity~\cite{peterson2018quantized, Liu2025Quantized}. Hence, it would be highly desirable to extend our system to two dimensions, where novel physics emerging from the interplay among nonlinearity, non-Hermiticity, and band topology can be further explored.}


\noindent\textbf{Methods}\\
\noindent\textit{Nonlinear eigenproblem.}

We employ a self-consistent method implemented in MATLAB to solve the nonlinear eigen equation ${H}(\psi)\psi=E\psi$ to obtain the eigenvalues and eigenmodes shown in Figs.~\ref{fig1} and  \ref{fig5}. For a one-dimensional chain with $N$ sites, the eigen problem involves $2N+2$ unknown real variables to be solved, including the real and imaginary parts of each component of the wavefunction (i.e., $\text{Re}(\psi_n)$ and $\text{Im}(\psi_n)$), and the real and imaginary parts of the eigenvalue (i.e.,  $\text{Re}(E)$ and $\text{Im}(E)$). Yet there are only $2N+1$ equations, including $2N$ equations from the eigen equation and the intensity condition $I=\sum_n |\psi_n|^2$. To solve this issue, consider a modification to the eigen problem, ${H}(\psi)e^{i\phi}\psi=E e^{i\phi}\psi$, which states that when $\psi$ is a solution, $e^{i\phi}\psi$ is also a solution. Thus, we can also find a proper $\phi$ to make the first component of the wavefunction be a real number (i.e., $\text{Im}(\psi_1)=0$), which reduces the unknown variables to $2N+1$. In our calculation, we start with a small intensity value $I=0.02$ and use the linear eigenvalues and eigenmodes as initial guesses to input into the solver. Then, the intensity is gradually increased ($\delta I=0.02$), with the initial guess for each step being the solutions from the previous step to ensure accuracy and computational efficiency.\\

\noindent\textit{Sample details.} The microwave resonators in this work are fabricated with printed circuit board (PCB) technology.  Both sides of the microstrip board are coated with 35-$\mu$m-thick copper layers, and the dielectric substrate is composed of 0.8-mm-thick FR4 material with a dielectric constant of 4.6 and a loss tangent of 0.02.  The resonator is H-shaped~\cite{peterson2018quantized}, where each section has the same length (2 cm), width (0.1 cm) and characteristic impedance (about 64 $\Omega$). The unloaded resonance frequency (1.2 GHz) is decreased by coupling capacitors. A resistor is patched in the middle of the resonator for system stability. The LNA chip (MGA-53543) is designed with a bias circuit while back-to-back varactors (SMV2201) are blocked with an isolated inductor (see Supplementary Note 4).\\

\noindent\textit{Measurement setup.} Measurements of  fundamental frequencies  are performed using a two-port vector network analyzer (VNA) (AV3672E) with tunable input power. In the measurement of the S-parameters (i.e., Fig.~\ref{fig2}), the source and detector are directly connected to the two SMA ports of the resonators via RF flexible coax cables. In the port measurement of the lattice sample (i.e., Fig.~\ref{fig3}b-d), the source signal is divided into 11 parts with identical magnitude by power dividers before input into the 11 resonators. Then the detector collects the transmission spectra one by one at the other port of each resonator. The measured frequency range is from 0.9 GHz to 1.3 GHz, with a total of 401 frequency points measured across this bandwidth. In the near-field measurement (i.e., Fig.~\ref{fig3}i-k), the detector probe is 1 mm above the sample and the scanning setup involves a 6-point grid in the $x$ direction and a 71-point grid in the $y$ direction, with a spacing of 5.6 mm between each point.

Measurements of harmonics (Fig.~\ref{fig4}) are performed using a signal generator (SSG5060X-V) with tunable output power and frequency and a signal analyzer (N9020A). A continuous-wave source connects to resonator 1, generating a sinusoidal signal with frequencies from 0.95 GHz to 1.05 GHz in 2 MHz steps, and varying power from -55 dBm to 25 dBm in 1 dBm steps. The output spectrum is recorded using a signal analyzer over the frequency range of 0.9–3.1 GHz with a frequency resolution of 2 MHz.  To avoid overloading and  maintain precise power measurements, we employ a 15 dB attenuator in front of the analyzer's input port.

\noindent\textbf{Data Availability}\\
The raw data used to generate the plots are available at Zenodo: \url{https://doi.org/10.5281/zenodo.20559726}~\cite{wu2026sourcedata}. Other data supporting this study's findings are available from the corresponding authors upon request.

\noindent\textbf{Code Availability}\\
The codes used to generate the plots are available in the same Zenodo record: \url{https://doi.org/10.5281/zenodo.20559726}~\cite{wu2026sourcedata}. Other codes supporting this study's findings are available from the corresponding authors upon request.

\bibliography{Nonlinear-ref}

@article{su1979solitons,
  title   = {Solitons in polyacetylene},
  author  = {Su, Wu-Pei and Schrieffer, John Robert and Heeger, Alan J},
  journal = {Phys. Rev. Lett.},
  volume  = {42},
  number  = {25},
  pages   = {1698},
  year    = {1979},
  doi     = {10.1103/PhysRevLett.42.1698}
}

@article{hatano1996localization,
  title   = {Localization transitions in non-{H}ermitian quantum mechanics},
  author  = {Hatano, Naomichi and Nelson, David R},
  journal = {Phys. Rev. Lett.},
  volume  = {77},
  number  = {3},
  pages   = {570},
  year    = {1996},
  doi     = {10.1103/PhysRevLett.77.570}
}

@article{wang2008reflection,
  title   = {Reflection-free one-way edge modes in a gyromagnetic photonic crystal},
  author  = {Wang, Zheng and Chong, YD and Joannopoulos, John D and Solja{\v{c}}i{\'c}, Marin},
  journal = {Phys. Rev. Lett.},
  volume  = {100},
  number  = {1},
  pages   = {013905},
  year    = {2008},
  doi     = {10.1103/PhysRevLett.100.013905}
}

@article{wang2009observation,
  title   = {Observation of unidirectional backscattering-immune topological electromagnetic states},
  author  = {Wang, Zheng and Chong, Yidong and Joannopoulos, John D and Solja{\v{c}}i{\'c}, Marin},
  journal = {Nature},
  volume  = {461},
  number  = {7265},
  pages   = {772--775},
  year    = {2009},
  doi     = {10.1038/nature08293}
}

@article{khanikaev2013photonic,
  title   = {Photonic topological insulators},
  author  = {Khanikaev, Alexander B and Hossein Mousavi, S and Tse, Wang-Kong and Kargarian, Mehdi and MacDonald, Allan H and Shvets, Gennady},
  journal = {Nat. Mater.},
  volume  = {12},
  number  = {3},
  pages   = {233--239},
  year    = {2013},
  doi     = {10.1038/nmat3520}
}

@article{liang2013optical,
  title   = {Optical resonator analog of a two-dimensional topological insulator},
  author  = {Liang, GQ and Chong, YD},
  journal = {Phys. Rev. Lett.},
  volume  = {110},
  number  = {20},
  pages   = {203904},
  year    = {2013},
  doi     = {10.1103/PhysRevLett.110.203904}
}

@article{lumer2013self,
  title   = {Self-localized states in photonic topological insulators},
  author  = {Lumer, Yaakov and Plotnik, Yonatan and Rechtsman, Mikael C and Segev, Mordechai},
  journal = {Phys. Rev. Lett.},
  volume  = {111},
  number  = {24},
  pages   = {243905},
  year    = {2013},
  doi     = {10.1103/PhysRevLett.111.243905}
}

@article{rechtsman2013photonic,
  title   = {Photonic {F}loquet topological insulators},
  author  = {Rechtsman, Mikael C and Zeuner, Julia M and Plotnik, Yonatan and Lumer, Yaakov and Podolsky, Daniel and Dreisow, Felix and Nolte, Stefan and Segev, Mordechai and Szameit, Alexander},
  journal = {Nature},
  volume  = {496},
  number  = {7444},
  pages   = {196--200},
  year    = {2013},
  doi     = {10.1038/nature12066}
}

@article{chen2014experimental,
  title   = {Experimental realization of photonic topological insulator in a uniaxial metacrystal waveguide},
  author  = {Chen, Wen-Jie and Jiang, Shao-Ji and Chen, Xiao-Dong and Zhu, Baocheng and Zhou, Lei and Dong, Jian-Wen and Chan, Che Ting},
  journal = {Nat. Commun.},
  volume  = {5},
  number  = {1},
  pages   = {5782},
  year    = {2014},
  doi     = {10.1038/ncomms6782}
}

@article{lu2014topological,
  title   = {Topological photonics},
  author  = {Lu, Ling and Joannopoulos, John D and Solja{\v{c}}i{\'c}, Marin},
  journal = {Nat. Photon.},
  volume  = {8},
  number  = {11},
  pages   = {821--829},
  year    = {2014},
  doi     = {10.1038/nphoton.2014.248}
}

@article{wu2015scheme,
  title   = {Scheme for achieving a topological photonic crystal by using dielectric material},
  author  = {Wu, Long-Hua and Hu, Xiao},
  journal = {Phys. Rev. Lett.},
  volume  = {114},
  number  = {22},
  pages   = {223901},
  year    = {2015},
  doi     = {10.1103/PhysRevLett.114.223901}
}

@article{gao2016probing,
  title   = {Probing topological protection using a designer surface plasmon structure},
  author  = {Gao, Fei and Gao, Zhen and Shi, Xihang and Yang, Zhaoju and Lin, Xiao and Xu, Hongyi and Joannopoulos, John D and Solja{\v{c}}i{\'c}, Marin and Chen, Hongsheng and Lu, Ling and others},
  journal = {Nat. Commun.},
  volume  = {7},
  number  = {1},
  pages   = {11619},
  year    = {2016},
  doi     = {10.1038/ncomms11619}
}

@article{hadad2016self,
  title   = {Self-induced topological transitions and edge states supported by nonlinear staggered potentials},
  author  = {Hadad, Yakir and Khanikaev, Alexander B and Alu, Andrea},
  journal = {Phys. Rev. B},
  volume  = {93},
  number  = {15},
  pages   = {155112},
  year    = {2016},
  doi     = {10.1103/PhysRevB.93.155112}
}

@article{leykam2016edge,
  title   = {Edge solitons in nonlinear-photonic topological insulators},
  author  = {Leykam, Daniel and Chong, Yi Dong},
  journal = {Phys. Rev. Lett.},
  volume  = {117},
  number  = {14},
  pages   = {143901},
  year    = {2016},
  doi     = {10.1103/PhysRevLett.117.143901}
}

@article{ma2016all,
  title   = {All-{S}i valley-{H}all photonic topological insulator},
  author  = {Ma, Tzuhsuan and Shvets, Gennady},
  journal = {New J. Phys.},
  volume  = {18},
  number  = {2},
  pages   = {025012},
  year    = {2016},
  doi     = {10.1088/1367-2630/18/2/025012}
}

@article{dong2017valley,
  title   = {Valley photonic crystals for control of spin and topology},
  author  = {Dong, Jian-Wen and Chen, Xiao-Dong and Zhu, Hanyu and Wang, Yuan and Zhang, Xiang},
  journal = {Nat. Mater.},
  volume  = {16},
  number  = {3},
  pages   = {298--302},
  year    = {2017},
  doi     = {10.1038/nmat4807}
}

@article{dobrykh2018nonlinear,
  title   = {Nonlinear control of electromagnetic topological edge states},
  author  = {Dobrykh, DA and Yulin, AV and Slobozhanyuk, AP and Poddubny, AN and Kivshar, Yu S},
  journal = {Phys. Rev. Lett.},
  volume  = {121},
  number  = {16},
  pages   = {163901},
  year    = {2018},
  doi     = {10.1103/PhysRevLett.121.163901}
}

@article{gao2018topologically,
  title   = {Topologically protected refraction of robust kink states in valley photonic crystals},
  author  = {Gao, Fei and Xue, Haoran and Yang, Zhaoju and Lai, Kueifu and Yu, Yang and Lin, Xiao and Chong, Yidong and Shvets, Gennady and Zhang, Baile},
  journal = {Nat. Phys.},
  volume  = {14},
  number  = {2},
  pages   = {140--144},
  year    = {2018},
  doi     = {10.1038/nphys4304}
}

@article{gong2018topological,
  title   = {Topological phases of non-{H}ermitian systems},
  author  = {Gong, Zongping and Ashida, Yuto and Kawabata, Kohei and Takasan, Kazuaki and Higashikawa, Sho and Ueda, Masahito},
  journal = {Phys. Rev. X},
  volume  = {8},
  number  = {3},
  pages   = {031079},
  year    = {2018},
  doi     = {10.1103/PhysRevX.8.031079}
}

@article{kunst2018biorthogonal,
  title   = {Biorthogonal bulk-boundary correspondence in non-{H}ermitian systems},
  author  = {Kunst, Flore K and Edvardsson, Elisabet and Budich, Jan Carl and Bergholtz, Emil J},
  journal = {Phys. Rev. Lett.},
  volume  = {121},
  number  = {2},
  pages   = {026808},
  year    = {2018},
  doi     = {10.1103/PhysRevLett.121.026808}
}

@article{leykam2018reconfigurable,
  title   = {Reconfigurable topological phases in next-nearest-neighbor coupled resonator lattices},
  author  = {Leykam, Daniel and Mittal, S and Hafezi, M and Chong, Yi Dong},
  journal = {Phys. Rev. Lett.},
  volume  = {121},
  number  = {2},
  pages   = {023901},
  year    = {2018},
  doi     = {10.1103/PhysRevLett.121.023901}
}

@article{longhi2018non,
  title   = {Non-{H}ermitian gauged topological laser arrays},
  author  = {Longhi, Stefano},
  journal = {Ann. Phys.},
  volume  = {530},
  number  = {7},
  pages   = {1800023},
  year    = {2018},
  doi     = {10.1002/andp.201800023}
}

@article{martinez2018non,
  title   = {Non-{H}ermitian robust edge states in one dimension: {A}nomalous localization and eigenspace condensation at exceptional points},
  author  = {Martinez Alvarez, VM and Barrios Vargas, JE and Foa Torres, LEF},
  journal = {Phys. Rev. B},
  volume  = {97},
  number  = {12},
  pages   = {121401},
  year    = {2018},
  doi     = {10.1103/PhysRevB.97.121401}
}

@article{peterson2018quantized,
  title   = {A quantized microwave quadrupole insulator with topologically protected corner states},
  author  = {Peterson, Christopher W and Benalcazar, Wladimir A and Hughes, Taylor L and Bahl, Gaurav},
  journal = {Nature},
  volume  = {555},
  number  = {7696},
  pages   = {346--350},
  year    = {2018},
  doi     = {10.1038/nature25777}
}

@article{yao2018edge,
  title   = {Edge states and topological invariants of non-{H}ermitian systems},
  author  = {Yao, Shunyu and Wang, Zhong},
  journal = {Phys. Rev. Lett.},
  volume  = {121},
  number  = {8},
  pages   = {086803},
  year    = {2018},
  doi     = {10.1103/PhysRevLett.121.086803}
}

@article{kawabata2019symmetry,
  title   = {Symmetry and topology in non-{H}ermitian physics},
  author  = {Kawabata, Kohei and Shiozaki, Ken and Ueda, Masahito and Sato, Masatoshi},
  journal = {Phys. Rev. X},
  volume  = {9},
  number  = {4},
  pages   = {041015},
  year    = {2019},
  doi     = {10.1103/PhysRevX.9.041015}
}

@article{kruk2019nonlinear,
  title   = {Nonlinear light generation in topological nanostructures},
  author  = {Kruk, Sergey and Poddubny, Alexander and Smirnova, Daria and Wang, Lei and Slobozhanyuk, Alexey and Shorokhov, Alexander and Kravchenko, Ivan and Luther-Davies, Barry and Kivshar, Yuri},
  journal = {Nat. Nanotechnol.},
  volume  = {14},
  number  = {2},
  pages   = {126--130},
  year    = {2019},
  doi     = {10.1038/s41565-018-0324-7}
}

@article{ozawa2019topological,
  title   = {Topological photonics},
  author  = {Ozawa, Tomoki and Price, Hannah M and Amo, Alberto and Goldman, Nathan and Hafezi, Mohammad and Lu, Ling and Rechtsman, Mikael C and Schuster, David and Simon, Jonathan and Zilberberg, Oded and others},
  journal = {Rev. Mod. Phys.},
  volume  = {91},
  number  = {1},
  pages   = {015006},
  year    = {2019},
  doi     = {10.1103/RevModPhys.91.015006}
}

@article{smirnova2019third,
  title   = {Third-harmonic generation in photonic topological metasurfaces},
  author  = {Smirnova, Daria and Kruk, Sergey and Leykam, Daniel and Melik-Gaykazyan, Elizaveta and Choi, Duk-Yong and Kivshar, Yuri},
  journal = {Phys. Rev. Lett.},
  volume  = {123},
  number  = {10},
  pages   = {103901},
  year    = {2019},
  doi     = {10.1103/PhysRevLett.123.103901}
}

@article{wang2019topologically,
  title   = {Topologically enhanced harmonic generation in a nonlinear transmission line metamaterial},
  author  = {Wang, You and Lang, Li-Jun and Lee, Ching Hua and Zhang, Baile and Chong, YD},
  journal = {Nat. Commun.},
  volume  = {10},
  number  = {1},
  pages   = {1102},
  year    = {2019},
  doi     = {10.1038/s41467-019-08966-9}
}

@article{zhou2019periodic,
  title   = {Periodic table for topological bands with non-{H}ermitian symmetries},
  author  = {Zhou, Hengyun and Lee, Jong Yeon},
  journal = {Phys. Rev. B},
  volume  = {99},
  number  = {23},
  pages   = {235112},
  year    = {2019},
  doi     = {10.1103/PhysRevB.99.235112}
}

@article{borgnia2020non,
  title   = {Non-{H}ermitian boundary modes and topology},
  author  = {Borgnia, Dan S and Kruchkov, Alex Jura and Slager, Robert-Jan},
  journal = {Phys. Rev. Lett.},
  volume  = {124},
  number  = {5},
  pages   = {056802},
  year    = {2020},
  doi     = {10.1103/PhysRevLett.124.056802}
}

@article{budich2020non,
  title   = {Non-{H}ermitian topological sensors},
  author  = {Budich, Jan Carl and Bergholtz, Emil J},
  journal = {Phys. Rev. Lett.},
  volume  = {125},
  number  = {18},
  pages   = {180403},
  year    = {2020},
  doi     = {10.1103/PhysRevLett.125.180403}
}

@article{lan2020nonlinear,
  title   = {Nonlinear one-way edge-mode interactions for frequency mixing in topological photonic crystals},
  author  = {Lan, Zhihao and You, Jian Wei and Panoiu, Nicolae C},
  journal = {Phys. Rev. B},
  volume  = {101},
  number  = {15},
  pages   = {155422},
  year    = {2020},
  doi     = {10.1103/PhysRevB.101.155422}
}

@article{maczewsky2020nonlinearity,
  title   = {Nonlinearity-induced photonic topological insulator},
  author  = {Maczewsky, Lukas J and Heinrich, Matthias and Kremer, Mark and Ivanov, Sergey K and Ehrhardt, Max and Martinez, Franklin and Kartashov, Yaroslav V and Konotop, Vladimir V and Torner, Lluis and Bauer, Dieter and others},
  journal = {Science},
  volume  = {370},
  number  = {6517},
  pages   = {701--704},
  year    = {2020},
  doi     = {10.1126/science.abd2033}
}

@article{mcdonald2020exponentially,
  title   = {Exponentially-enhanced quantum sensing with non-{H}ermitian lattice dynamics},
  author  = {McDonald, Alexander and Clerk, Aashish A},
  journal = {Nat. Commun.},
  volume  = {11},
  number  = {1},
  pages   = {5382},
  year    = {2020},
  doi     = {10.1038/s41467-020-19090-4}
}

@article{mukherjee2020observation,
  title   = {Observation of {F}loquet solitons in a topological bandgap},
  author  = {Mukherjee, Sebabrata and Rechtsman, Mikael C},
  journal = {Science},
  volume  = {368},
  number  = {6493},
  pages   = {856--859},
  year    = {2020},
  doi     = {10.1126/science.aba8725}
}

@article{okuma2020topological,
  title   = {Topological origin of non-{H}ermitian skin effects},
  author  = {Okuma, Nobuyuki and Kawabata, Kohei and Shiozaki, Ken and Sato, Masatoshi},
  journal = {Phys. Rev. Lett.},
  volume  = {124},
  number  = {8},
  pages   = {086801},
  year    = {2020},
  doi     = {10.1103/PhysRevLett.124.086801}
}

@article{smirnova2020nonlinear,
  title   = {Nonlinear topological photonics},
  author  = {Smirnova, Daria and Leykam, Daniel and Chong, Yidong and Kivshar, Yuri},
  journal = {Appl. Phys. Rev.},
  volume  = {7},
  number  = {2},
  pages   = {021306},
  year    = {2020},
  doi     = {10.1063/1.5142397}
}

@article{weidemann2020topological,
  title   = {Topological funneling of light},
  author  = {Weidemann, Sebastian and Kremer, Mark and Helbig, Tobias and Hofmann, Tobias and Stegmaier, Alexander and Greiter, Martin and Thomale, Ronny and Szameit, Alexander},
  journal = {Science},
  volume  = {368},
  number  = {6488},
  pages   = {311--314},
  year    = {2020},
  doi     = {10.1126/science.aaz8727}
}

@article{xiao2020non,
  title   = {Non-{H}ermitian bulk--boundary correspondence in quantum dynamics},
  author  = {Xiao, Lei and Deng, Tianshu and Wang, Kunkun and Zhu, Gaoyan and Wang, Zhong and Yi, Wei and Xue, Peng},
  journal = {Nat. Phys.},
  volume  = {16},
  number  = {7},
  pages   = {761--766},
  year    = {2020},
  doi     = {10.1038/s41567-020-0836-6}
}

@article{you2020four,
  title   = {Four-wave mixing of topological edge plasmons in graphene metasurfaces},
  author  = {You, Jian Wei and Lan, Zhihao and Panoiu, Nicolae C},
  journal = {Sci. Adv.},
  volume  = {6},
  number  = {13},
  pages   = {eaaz3910},
  year    = {2020},
  doi     = {10.1126/sciadv.aaz3910}
}

@article{zhang2020correspondence,
  title   = {Correspondence between winding numbers and skin modes in non-{H}ermitian systems},
  author  = {Zhang, Kai and Yang, Zhesen and Fang, Chen},
  journal = {Phys. Rev. Lett.},
  volume  = {125},
  number  = {12},
  pages   = {126402},
  year    = {2020},
  doi     = {10.1103/PhysRevLett.125.126402}
}

@article{bergholtz2021exceptional,
  title   = {Exceptional topology of non-{H}ermitian systems},
  author  = {Bergholtz, Emil J and Budich, Jan Carl and Kunst, Flore K},
  journal = {Rev. Mod. Phys.},
  volume  = {93},
  number  = {1},
  pages   = {015005},
  year    = {2021},
  doi     = {10.1103/RevModPhys.93.015005}
}

@article{hu2021nonlinear,
  title   = {Nonlinear control of photonic higher-order topological bound states in the continuum},
  author  = {Hu, Zhichan and Bongiovanni, Domenico and Juki{\'c}, Dario and Jajti{\'c}, Ema and Xia, Shiqi and Song, Daohong and Xu, Jingjun and Morandotti, Roberto and Buljan, Hrvoje and Chen, Zhigang},
  journal = {Light Sci. Appl.},
  volume  = {10},
  number  = {1},
  pages   = {164},
  year    = {2021},
  doi     = {10.1038/s41377-021-00607-5}
}

@article{jurgensen2021quantized,
  title   = {Quantized nonlinear {T}houless pumping},
  author  = {J{\"u}rgensen, Marius and Mukherjee, Sebabrata and Rechtsman, Mikael C},
  journal = {Nature},
  volume  = {596},
  number  = {7870},
  pages   = {63--67},
  year    = {2021},
  doi     = {10.1038/s41586-021-03688-9}
}

@article{kirsch2021nonlinear,
  title   = {Nonlinear second-order photonic topological insulators},
  author  = {Kirsch, Marco S and Zhang, Yiqi and Kremer, Mark and Maczewsky, Lukas J and Ivanov, Sergey K and Kartashov, Yaroslav V and Torner, Lluis and Bauer, Dieter and Szameit, Alexander and Heinrich, Matthias},
  journal = {Nat. Phys.},
  volume  = {17},
  number  = {9},
  pages   = {995--1000},
  year    = {2021},
  doi     = {10.1038/s41567-021-01275-3}
}

@article{xia2021nonlinear,
  title   = {Nonlinear tuning of {PT} symmetry and non-{H}ermitian topological states},
  author  = {Xia, Shiqi and Kaltsas, Dimitrios and Song, Daohong and Komis, Ioannis and Xu, Jingjun and Szameit, Alexander and Buljan, Hrvoje and Makris, Konstantinos G and Chen, Zhigang},
  journal = {Science},
  volume  = {372},
  number  = {6537},
  pages   = {72--76},
  year    = {2021},
  doi     = {10.1126/science.abf6873}
}

@article{yuce2021nonlinear,
  title   = {Nonlinear non-{H}ermitian skin effect},
  author  = {Yuce, Cem},
  journal = {Phys. Lett. A},
  volume  = {408},
  pages   = {127484},
  year    = {2021},
  doi     = {10.1016/j.physleta.2021.127484}
}

@article{zhu2021delocalization,
  title   = {Delocalization of topological edge states},
  author  = {Zhu, Weiwei and Teo, Wei Xin and Li, Linhu and Gong, Jiangbin},
  journal = {Phys. Rev. B},
  volume  = {103},
  number  = {19},
  pages   = {195414},
  year    = {2021},
  doi     = {10.1103/PhysRevB.103.195414}
}

@article{ding2022non,
  title   = {Non-{H}ermitian topology and exceptional-point geometries},
  author  = {Ding, Kun and Fang, Chen and Ma, Guancong},
  journal = {Nat. Rev. Phys.},
  volume  = {4},
  number  = {12},
  pages   = {745--760},
  year    = {2022},
  doi     = {10.1038/s42254-022-00516-5}
}

@article{liu2022complex,
  title   = {Complex skin modes in non-{H}ermitian coupled laser arrays},
  author  = {Liu, Yuzhou GN and Wei, Yunxuan and Hemmatyar, Omid and Pyrialakos, Georgios G and Jung, Pawel S and Christodoulides, Demetrios N and Khajavikhan, Mercedeh},
  journal = {Light Sci. Appl.},
  volume  = {11},
  number  = {1},
  pages   = {336},
  year    = {2022},
  doi     = {10.1038/s41377-022-01030-0}
}

@article{wang2022non,
  title   = {Non-{H}ermitian morphing of topological modes},
  author  = {Wang, Wei and Wang, Xulong and Ma, Guancong},
  journal = {Nature},
  volume  = {608},
  number  = {7921},
  pages   = {50--55},
  year    = {2022},
  doi     = {10.1038/s41586-022-04929-1}
}

@article{ezawa2022dynamical,
  title   = {Dynamical nonlinear higher-order non-{H}ermitian skin effects and topological trap-skin phase},
  author  = {Ezawa, Motohiko},
  journal = {Phys. Rev. B},
  volume  = {105},
  number  = {12},
  pages   = {125421},
  year    = {2022},
  doi     = {10.1103/PhysRevB.105.125421}
}

@article{zhu2022anomalous,
  title   = {Anomalous single-mode lasing induced by nonlinearity and the non-{H}ermitian skin effect},
  author  = {Zhu, Bofeng and Wang, Qiang and Leykam, Daniel and Xue, Haoran and Wang, Qi Jie and Chong, YD},
  journal = {Phys. Rev. Lett.},
  volume  = {129},
  number  = {1},
  pages   = {013903},
  year    = {2022},
  doi     = {10.1103/PhysRevLett.129.013903}
}

@article{manna2023inner,
  title   = {Inner skin effects on non-{H}ermitian topological fractals},
  author  = {Manna, Sourav and Roy, Bitan},
  journal = {Commun. Phys.},
  volume  = {6},
  number  = {1},
  pages   = {10},
  year    = {2023},
  doi={https://doi.org/10.1038/s42005-023-01130-2}
}

@article{gao2023two,
  title   = {Two-dimensional reconfigurable non-{H}ermitian gauged laser array},
  author  = {Gao, Zihe and Qiao, Xingdu and Pan, Mingsen and Wu, Shuang and Yim, Jieun and Chen, Kaiyuan and Midya, Bikashkali and Ge, Li and Feng, Liang},
  journal = {Phys. Rev. Lett.},
  volume  = {130},
  number  = {26},
  pages   = {263801},
  year    = {2023},
  doi     = {10.1103/PhysRevLett.130.263801}
}

@article{jiang2023nonlinear,
  title   = {Nonlinear perturbation of a high-order exceptional point: Skin discrete breathers and the hierarchical power-law scaling},
  author  = {Jiang, Hui and Cheng, Enhong and Zhou, Ziyu and Lang, Li-Jun},
  journal = {Chin. Phys. B},
  volume  = {32},
  number  = {8},
  pages   = {084203},
  year    = {2023},
  doi     = {10.1088/1674-1056/accb47}
}

@article{dai2024non,
  title   = {Non-{H}ermitian topological phase transitions controlled by nonlinearity},
  author  = {Dai, Tianxiang and Ao, Yutian and Mao, Jun and Yang, Yan and Zheng, Yun and Zhai, Chonghao and Li, Yandong and Yuan, Jingze and Tang, Bo and Li, Zhihua and others},
  journal = {Nat. Phys.},
  volume  = {20},
  number  = {1},
  pages   = {101--108},
  year    = {2024},
  doi     = {10.1038/s41567-023-02244-8}
}

@article{deng2024ultrasensitive,
  title   = {Ultrasensitive integrated circuit sensors based on high-order non-{H}ermitian topological physics},
  author  = {Deng, Wenyuan and Zhu, Wei and Chen, Tian and Sun, Houjun and Zhang, Xiangdong},
  journal = {Sci. Adv.},
  volume  = {10},
  number  = {38},
  pages   = {eadp6905},
  year    = {2024},
  doi     = {10.1126/sciadv.adp6905}
}

@article{ghaemi2024class,
  title   = {A class of stable nonlinear non-{H}ermitian skin modes},
  author  = {Ghaemi-Dizicheh, Hamed},
  journal = {Phys. Scr.},
  volume  = {99},
  number  = {12},
  pages   = {125411},
  year    = {2024},
  doi     = {10.1088/1402-4896/ad91f0}
}

@article{liu2024localization,
  title   = {Localization of chiral edge states by the non-{H}ermitian skin effect},
  author  = {Liu, Gui-Geng and Mandal, Subhaskar and Zhou, Peiheng and Xi, Xiang and Banerjee, Rimi and Hu, Yuan-Hang and Wei, Minggui and Wang, Maoren and Wang, Qiang and Gao, Zhen and others},
  journal = {Phys. Rev. Lett.},
  volume  = {132},
  number  = {11},
  pages   = {113802},
  year    = {2024},
  doi     = {10.1103/PhysRevLett.132.113802}
}

@article{lo2024switchable,
  title         = {Switchable non-{H}ermitian skin effect in {B}ogoliubov modes},
  author        = {Lo, Hsuan and Wang, You and Banerjee, Rimi and Zhang, Baile and Chong, YD},
  journal       = {arXiv},
  eprint        = {2411.13841},
  archivePrefix = {arXiv},
  year          = {2024},
  url           = {https://arxiv.org/abs/2411.13841}
}

@article{many2024skin,
  title   = {Skin modes in a nonlinear {H}atano-{N}elson model},
  author  = {Many Manda, Bertin and Carretero-Gonz{\'a}lez, Ricardo and Kevrekidis, Panayotis G and Achilleos, Vassos},
  journal = {Phys. Rev. B},
  volume  = {109},
  number  = {9},
  pages   = {094308},
  year    = {2024},
  doi     = {10.1103/PhysRevB.109.094308}
}

@article{sun2024photonic,
  title   = {Photonic {F}loquet skin-topological effect},
  author  = {Sun, Yeyang and Hou, Xiangrui and Wan, Tuo and Wang, Fangyu and Zhu, Shiyao and Ruan, Zhichao and Yang, Zhaoju},
  journal = {Phys. Rev. Lett.},
  volume  = {132},
  number  = {6},
  pages   = {063804},
  year    = {2024},
  doi     = {10.1103/PhysRevLett.132.063804}
}

@article{wang2024nonlinear,
  title         = {Nonlinear non-{H}ermitian skin effect and skin solitons in temporal photonic feedback lattices},
  author        = {Wang, Shulin and Wang, Bing and Liu, Chenyu and Qin, Chengzhi and Zhao, Lange and Liu, Weiwei and Longhi, Stefano and Lu, Peixiang},
  journal       = {arXiv},
  eprint        = {2409.19693},
  archivePrefix = {arXiv},
  year          = {2024},
  url           = {https://arxiv.org/abs/2409.19693}
}

@article{padlewski2025observation,
  title   = {Observation of amplitude-driven nonreciprocity for energy guiding},
  author  = {Padlewski, Mathieu and Fleury, Romain and Lissek, Herv{\'e}},
  journal = {Phys. Rev. B},
  volume  = {111},
  number  = {12},
  pages   = {125156},
  year    = {2025},
  doi     = {10.1103/PhysRevB.111.125156}
}

@article{yuce2025nonlinear,
  title   = {Nonlinear skin modes and fixed points},
  author  = {Yuce, C},
  journal = {Phys. Rev. B},
  volume  = {111},
  number  = {5},
  pages   = {054201},
  year    = {2025},
  doi     = {10.1103/PhysRevB.111.054201}
}

@article{sahin2025topolectrical,
  title   = {Topolectrical circuits---recent experimental advances and developments},
  author  = {Sahin, Haydar and Jalil, Mansoor and Lee, Ching Hua},
  journal = {APL Electron. Devices},
  volume  = {1},
  number  = {2},
  year    = {2025}
}

@article{Liu2025Quantized,
  title   = {Quantized decay charges in non-{H}ermitian networks characterized by directed graphs},
  author  = {Liu, Wenwen and Wu, Junyao and Zhang, Li and You, Oubo and Tian, Ye and Chen, Hongsheng and Min, Bumki and Yang, Yihao and Zhang, Shuang},
  journal = {Phys. Rev. Lett.},
  volume  = {135},
  number  = {20},
  pages   = {206602},
  year    = {2025},
  doi     = {10.1103/dqf4-6fg5},
  url     = {https://link.aps.org/doi/10.1103/dqf4-6fg5}
}

@article{sahin2025protected,
  title   = {Protected chaos in a topological lattice},
  author  = {Sahin, Haydar and Akg{\"u}n, Hakan and Siu, Zhuo Bin and Rafi-Ul-Islam, SM and Kong, Jian Feng and Jalil, Mansoor BA and Lee, Ching Hua},
  journal = {Adv. Sci.},
  volume  = {12},
  number  = {28},
  pages   = {e03216},
  year={2025},
  publisher={Wiley Online Library},
  doi     = { https://doi.org/10.1002/advs.202503216}
}

@misc{wu2026sourcedata,
author = {Wu, Junyao and Shen, Rui-Chang and Zhang, Li and Chen, Fujia and Wang, Bingbing and Chen, Hongsheng and Yang, Yihao and Xue, Haoran},
title = {Source data and code for ``Observation of localization reversal and harmonic generation in nonlinear non-Hermitian skin effect''},
year = {2026},
publisher = {Zenodo},
doi = {10.5281/zenodo.20559726},
url = {https://doi.org/10.5281/zenodo.20559726}
}

\noindent\textbf{Acknowledgements}\\
We are grateful to Linhu Li for helpful discussions.\\
\noindent\textbf{Funding}\\
The work at Zhejiang University was sponsored by the Key Research and Development Program of the Ministry of Science and Technology under Grants No.~2022YFA1405200 (Y.Y.), No.~2022YFA1404900 (Y.Y.), No.~2022YFA1404704 (H.C.), and No.~2022YFA1404902 (H.C.), the National Natural Science Foundation of China (NNSFC) under Grants No.~62175215 (Y.Y.), No.~61975176 (H.C.), the Key Research and Development Program of Zhejiang Province under Grant No.~2022C01036 (H.C.), the Fundamental Research Funds for the Central Universities (No.~2021FZZX001-19) (Y.Y.), the Excellent Young Scientists Fund Program (Overseas) of China (Y.Y.). The work at Chinese University of Hong Kong was supported by the National Natural Science Foundation of China under Grant No.~62401491, the Research Grants Council of the Hong Kong Special Administrative Region, China, under Grant No.~24304825, and the Chinese University of Hong Kong under Grants No.~4937205, No.~4937206, No.~4053729 and No.~4411765.\\
\noindent\textbf{Author contributions}\\
H.X., Y.Y. and H.C. conceived the idea. J.W. designed the experimental structure and carried out the experiments. L.Z. and F.C. provided theoretical guidance throughout the study. R.-C.S. and B.W. performed the theoretical analysis. J.W., R.-C.S., H.C., Y.Y. and H.X. contributed to the writing of the manuscript and interpretation of the results. Y.Y. and H.X. supervised the project. All authors discussed the results and reviewed the manuscript.\\
\noindent\textbf{Competing interests}\\
The authors declare no competing interests.
\end{document}